\documentclass[letterpaper, 10 pt, conference]{ieeeconf}  

\makeatletter
\let\NAT@parse\undefined
\makeatother
\IEEEoverridecommandlockouts                            
\usepackage{graphicx}   
\usepackage{comment} 
\usepackage{amssymb,amsmath, mathrsfs,mathtools}
\usepackage{tabularray}
\usepackage{threeparttable}
\usepackage{multirow}
\usepackage{array}
\usepackage{bigints}
\usepackage{xurl}
\usepackage{cite}
\usepackage[colorlinks,allcolors=blue]{hyperref}

\title{\LARGE \bf Probabilistically Robust Uncertainty Analysis and Optimal Control of Continuous Lyophilization via Polynomial Chaos Theory}

\author{Prakitr Srisuma, George Barbastathis, and Richard D. Braatz 
\thanks{The authors are with the Massachusetts Institute of Technology, Cambridge, MA 02139. Email: \{prakitrs, gbarb, braatz\}@mit.edu}%
}

\begin{document}

\maketitle
\thispagestyle{empty}
\pagestyle{empty}

\def\thefootnote{}\footnotetext{\copyright \ 2025 IEEE. Personal use of this material is permitted. Permission from IEEE must be obtained for all other uses, in any current or future media, including reprinting/republishing this material for advertising or promotional purposes, creating new collective works, for resale or redistribution to servers or lists, or reuse of any copyrighted component of this work in other works.}
\begin{abstract}
Lyophilization, aka freeze drying, is a process commonly used to increase the stability of various drug products in biotherapeutics manufacturing, e.g., mRNA vaccines, allowing for higher storage temperature. While the current trends in the industry are moving towards continuous manufacturing, the majority of industrial lyophilization processes are still being operated in a batch mode. This article presents a framework that accounts for the probabilistic uncertainty during the primary and secondary drying steps in continuous lyophilization. The probabilistic uncertainty is incorporated into the mechanistic model via polynomial chaos theory (PCT). The resulting PCT-based model is able to accurately and efficiently quantify the effects of uncertainty on several critical process variables, including the temperature, sublimation front, and concentration of bound water. The integration of the PCT-based model into stochastic optimization and control is demonstrated. The proposed framework and case studies can be used to guide the design and control of continuous lyophilization while accounting for probabilistic uncertainty. 

\end{abstract}

\section{Introduction} \label{Introduction}

Lyophilization, aka freeze drying, is a process typically used to increase the stability of drug products in biotherapeutics manufacturing \cite{Fissore2018Review}. The process comprises three main steps, namely (1) freezing, (2) primary drying, and (3) secondary drying. During freezing, the product is cooled such that most of the liquid (free water) is frozen, with the remaining part (bound water) retaining its liquid state in the solid material. In primary drying, the temperature and pressure are reduced below the triple point, and so the free water in the form of ice crystals is removed via sublimation. In secondary drying, the temperature is increased further to reduce the amount of bound water via desorption, which is the final step of lyophilization. Various applications of lyophilization have been demonstrated, including for mRNA vaccines (e.g., for COVID-19), to allow for higher storage temperature \cite{Muramatsu2022mRNA,Meulewaeter2023mRNA}.

While the current trends in the industry are moving towards continuous manufacturing, the majority of production-scale lyophilization is still operated in batch mode \cite{Meyer2015SpinFreezing,Pisano2019ReviewContLyo}. A variety of continuous lyophilization concepts have been proposed (see a comprehensive review of continuous lyophilization technologies in \cite{Pisano2019ReviewContLyo}), with the most recent technology relying on the suspended-vial configuration \cite{Capozzi2019ContLyo_SuspendedVials}. In such cases, vials are suspended and continuously move along the process without any complicated motions (e.g., as in spin or spray freeze drying), allowing for product quality control that is difficult to achieve in alternative continuous lyophilization technology. 

Like all process models, lyophilization models have several associated uncertainties in the model parameters. The design and control of continuous lyophilization should be formulated so as to efficiently account for these uncertainties. However, previous studies mainly focused on batch lyophilization and analyzed only the primary drying step, in which the uncertainty was commonly considered for the heat transfer coefficient and cake resistance \cite{Giordano2011Model,Bosca2015Risk,Mortier2016Uncertainty,VanBockstal2017Risk,Bano2020LumpedDrying,Ravnik2021UQ}. To the authors' knowledge, no literature has yet considered the effects of uncertainties during continuous lyophilization and addressed these uncertainties in the control system design.

This article presents a framework that efficiently accounts for the probabilistic uncertainty during primary and secondary drying in continuous lyophilization. The state-of-the-art continuous lyophilization technique, via suspended vials, is considered, with polynomial chaos theory (PCT) used for probabilistically robust uncertainty analysis. Our main contributions are to
\begin{enumerate}
\item develop a PCT-based model for the continuous lyophilization of suspended vials,
\item employ the PCT-based model for efficient uncertainty quantification, process optimization, and control associated with probabilistic constraints, and
\item benchmark and validate the approach against the Monte Carlo method.
\end{enumerate}

This article is organized as follows. Section \ref{sec:ProcessAndModel} describes the process and corresponding mechanistic model. Section \ref{sec:PCE} discusses the mathematical formulation and implementation of PCT. Section \ref{sec:Results} presents various case studies on uncertainty quantification and stochastic optimization/control. Finally, Section \ref{sec:Conclusion} summarizes the study.


\section{Process Description and Modeling} \label{sec:ProcessAndModel}
This section describes the continuous lyophilization process considered in this study and summarizes the corresponding mechanistic models used in the later sections.

\subsection{Continuous lyophilization of suspended vials}
\begin{figure}[ht!]
\centering
\vspace{5pt}
\includegraphics[scale=.42]{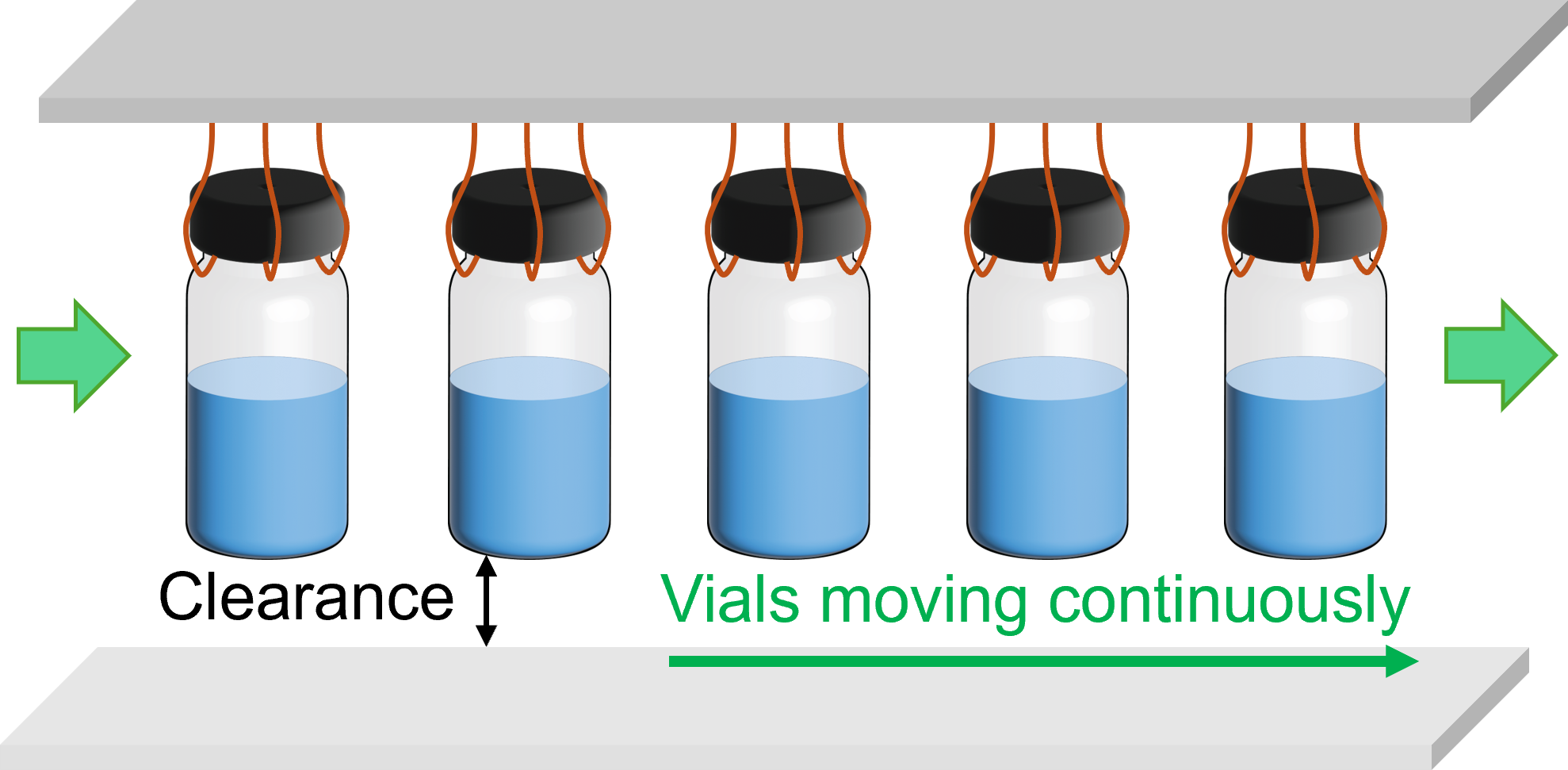}
    \vspace{-5pt}
    \caption{Continuous lyophilization of suspended vials. Vials are suspended and move continuously along the lyophilizer.} 
    \label{fig:SuspendedVials} 
    \vspace{-12pt} 
\end{figure}

This article considers a recently proposed state-of-the-art continuous lyophilization technology \cite{Capozzi2019ContLyo_SuspendedVials} that employs suspended vials continuously moving through the freezing and drying chambers, without any contact between the vials and shelf (Fig.~\ref{fig:SuspendedVials}). This suspended-vial lyophilization has various advantages, some of which are not applicable in existing technologies; see \cite{Capozzi2019ContLyo_SuspendedVials} for detailed information and discussion related to the suspended-vial lyophilization and \cite{Pisano2019ReviewContLyo} for a comprehensive review of continuous lyophilization technology in general.

\subsection{Mechanistic modeling} \label{sec:Model}
Mechanistic modeling for batch lyophilization has been studied for many decades (see examples and reviews in \cite{Fissore2015Review,Srisuma2024ContModeling}). Models for continuous lyophilization, however, are very limited, with most studies focusing spin freezing and spray freeze drying \cite{Sebastiao2019SprayFD,Nuytten2021SpinFreezing}. For the suspended-vial case, the only available model was recently developed \cite{Srisuma2024ContModeling}. Our mechanistic model is mainly based on \cite{Srisuma2024ContModeling}, which is summarized for the primary drying step in Section \ref{sec:Model1stDrying} and for the secondary drying step in Section \ref{sec:Model2ndDrying}. The freezing step is not considered in this work. 

\vspace{2 pt}
\subsubsection{Primary drying model} \label{sec:Model1stDrying}
\begin{figure}[ht!]
\centering
    \vspace{-6pt} 
    \includegraphics[scale=.8]{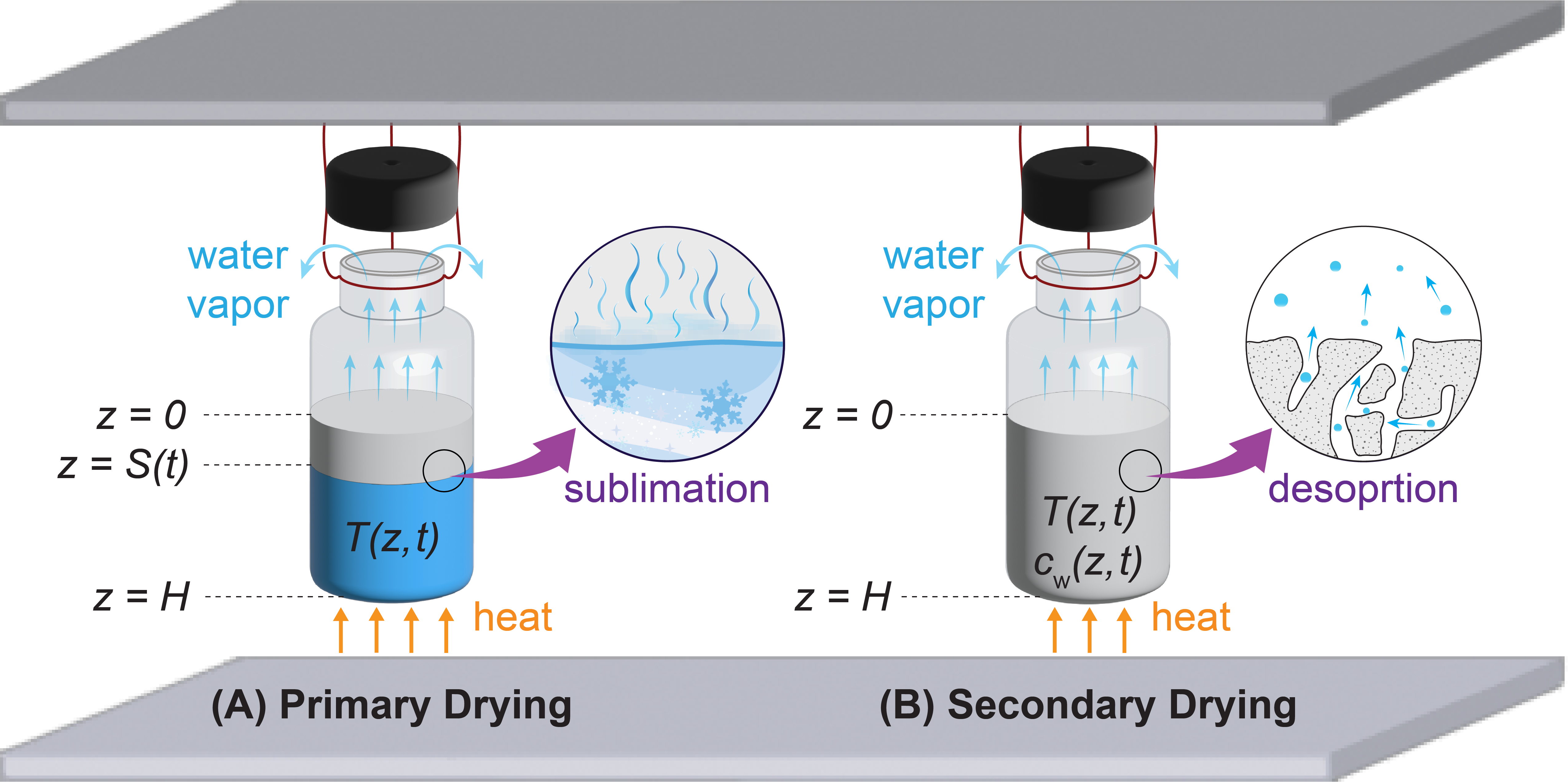}
    \vspace{-18pt} 
    \caption{Schematic diagram showing the mechanistic modeling of suspended vials for (A) primary drying and (B) secondary drying.} 
    \vspace{-7pt} 
    \label{fig:Schematic}      
\end{figure}

The model for primary drying is formulated in the rectangular coordinate system with one spatial dimension ($z$) and time ($t$) (Fig.~\ref{fig:Schematic}A). 

By assuming that the supplied heat mainly goes to the frozen region, the energy balance for the frozen region is
\begin{align}\label{eq:1st_energy}
\begin{split}
      \rho_\textrm{f} C_{p,\textrm{f}}\dfrac{\partial T}{\partial t} =  k_\textrm{f}\dfrac{\partial^2 T}{\partial z^2} + \frac{Q_\textrm{rad}}{\pi d^2 (H-S)},  \quad  S < z < H,
\end{split}
\end{align} 
where $T(z,t)$ is the temperature, $S(t)$ is the sublimation front/interface position, $H$ is the height of the product, $d$ is the vial diameter, $k$ is the thermal conductivity, $\rho$ is the density, and $C_p$ is the heat capacity, with the subscript `f' denoting the frozen region. The radiative heat transfer from the sidewall $Q_\textrm{rad}$ is
\begin{equation} \label{eq:1st_rad}
    Q_\textrm{rad} = \sigma A_r \mathcal{F}_1 (T_\textrm{c}^4-T^4),
\end{equation}
where $\mathcal{F}_1$ is the transfer factor, $\sigma$ is the Stefan-Boltzmann constant, $T_\textrm{c}$ is the chamber wall temperature, and $A_r=\pi dH$ is the side area of the product. At the sublimation front, the mass balance of water vapor is
\begin{equation}\label{eq:1st_interface}
    \frac{dS}{dt} = \frac{N_\textrm{w}}{\rho_\textrm{f}-\rho_\textrm{e}},
\end{equation}
where $N_\textrm{w}$ is the sublimation flux and $\rho_\textrm{e}$ is the effective density of the dried region (region above the sublimation front). The driving force for mass transfer is \cite{Pikal2005Model,Fissore2018Review,Bano2020LumpedDrying} 
\begin{equation}\label{eq:1st_flux}
    N_\textrm{w} = \frac{p_\textrm{w,sat}-p_\textrm{w,c}}{R_\textrm{p}},
\end{equation}
where $p_\textrm{w,sat}$ is the saturation/equilibrium pressure of water, $p_\textrm{w,c}$ is the partial pressure of water in the chamber (environment), and $R_\textrm{p}$ is the mass transfer resistance. The variation of the resistance can be approximated by the empirical expression \cite{Pikal2005Model,Fissore2015Review,Bano2020LumpedDrying}
\begin{equation}\label{eq:1st_Rp}
    R_\textrm{p} = R_0+ \frac{R_1S}{R_2+S},
\end{equation}
where $R_0$, $R_1$, and $R_2$ are the constants estimated from data. The saturation pressure is described by \cite{Bano2020LumpedDrying}
\begin{equation}\label{eq:1st_satpressure}
    p_\textrm{w,sat} =  \exp\!\left(\frac{-6139.9}{T} + 28.8912\right)\!.
\end{equation}
Natural convection and thermal radiation are the key heat transfer mechanisms for the suspended vials. The bottom surface of the frozen product is heated via natural convection and thermal radiation, which follows Newton's law of cooling
\begin{equation}\label{eq:1st_bcbottom}
    - k_\textrm{f}\dfrac{\partial T}{\partial z} = h(T-T_\textrm{b}),  \quad z=H, 
\end{equation}
where $T_\textrm{b}$ is the temperature of the bottom shelf and $h$ is the overall heat transfer coefficient that combines the effects of thermal radiation and natural convection. At the top surface, the energy balance associated with sublimation and thermal radiation is
\begin{equation}\label{eq:1st_bctop}
    N_\textrm{w}\Delta H_\textrm{sub}  =  k_\textrm{f}\dfrac{\partial T}{\partial z} + \sigma\mathcal{F}_2\left(T_\textrm{u}^4-T^4\right), \quad z=S,
\end{equation}
where $T_\textrm{u}$ is the temperature of the upper plate, $\Delta H_\textrm{sub}$ is the heat of sublimation, and $\mathcal{F}_2$ is the transfer factor. The initial conditions for \eqref{eq:1st_energy} and \eqref{eq:1st_interface} are
\begin{gather}
    T(z,t_0) = T_0, \quad 0\leq z \leq H, \label{eq:1st_iniT}  \\
    S(t_0) =  0, \label{eq:1st_iniS} 
\end{gather}
where $t_0$ is the initial time normally set to 0.

\vspace{2 pt}
\subsubsection{Secondary drying model} \label{sec:Model2ndDrying}
The model for secondary drying is formulated in the rectangular coordinate system with one spatial dimension ($z$) and time ($t$) (Fig.~\ref{fig:Schematic}B).

The energy balance of the dried product is 
\begin{align} \label{eq:2nd_energy}
\begin{split}
    \rho_\textrm{e} C_{p,\textrm{e}}\dfrac{\partial T}{\partial t} = k_\textrm{e}\dfrac{\partial^2 T}{\partial z^2}  + \rho_\textrm{d}\Delta H_\textrm{des}\dfrac{\partial c_\textrm{w}}{\partial t}  + \frac{Q_\textrm{rad}}{\pi d^2H},&  \\ \quad   0 \leq z \leq H,&
\end{split}
\end{align}
where $c_\textrm{w}(z,t)$ is the concentration of (aka residual water/moisture), $\rho_\textrm{d}$ is the density of the dried region (solid and vacuum), $\Delta H_\textrm{des}$ is the heat of desorption, $Q_\textrm{rad}$ is as defined in \eqref{eq:1st_rad}, and the other parameters are as defined in \eqref{eq:1st_energy}, with the subscript `e' denoting the effective properties considering both solid and gas in the pores. The desorption rate is described by the linear driving force model
\begin{equation} \label{eq:2nd_desorption}
     \dfrac{\partial c_\textrm{w}}{\partial t} = k_\textrm{d}(c^*_\textrm{w}-c_\textrm{w}),
\end{equation}
where $c^*_\textrm{w}$ is the equilibrium concentration and $k_\textrm{d}$ is the rate constant for desorption that exhibits Arrhenius temperature dependence \cite{Sadikoglu1997Modeling,Fissore2015Review} 
\begin{equation} \label{eq:2nd_kd}
     k_\textrm{d} = f_\textrm{a}e^{-E_\textrm{a}/RT},
\end{equation}
where $f_\textrm{a}$ is the frequency factor, $E_\textrm{a}$ is the activation energy, and $R$ is the gas constant. It is very common in the literature to simplify \eqref{eq:2nd_desorption} by setting $c^*_\textrm{w}=0$, which produces insignificant error \cite{Sadikoglu1997Modeling} and eliminates the need for equilibrium data and detailed knowledge about the solid structure \cite{Fissore2015Review}. The boundary condition at the bottom is similar to \eqref{eq:1st_bcbottom}, that is,
\begin{equation} \label{eq:2nd_bctbottom}
    - k_\textrm{e}\dfrac{\partial T}{\partial z} = h(T-T_\textrm{b}),  \quad z=H. 
\end{equation}
Heat transfer at the top surface is mainly thermal radiation, resulting in the boundary condition
\begin{equation} \label{eq:2nd_bctop}
     -k_\textrm{e}\dfrac{\partial T}{\partial z} =  \sigma\mathcal{F}_2 (T_\textrm{u}^4-T^4), \quad  z=0.
\end{equation}
The initial conditions for \eqref{eq:2nd_energy} and \eqref{eq:2nd_desorption} are
\begin{gather} 
    T(z,t_0) = T_0, \quad 0\leq z \leq H, \label{eq:2nd_iniT}  \\
    c_\textrm{w}(z,t_0) = c_\textrm{w,0},  \quad  0 \leq z \leq H. \label{eq:2nd_inicw} 
\end{gather}

The partial differential equations (PDEs) derived in Sections \ref{sec:Model1stDrying} and \ref{sec:Model2ndDrying} can be solved efficiently using the method of lines. We refer to \cite{Srisuma2024ContModeling} for the details of all relevant numerical methods. 


\section{Polynomial Chaos Theory} \label{sec:PCE}
Various techniques have been used to study and account for probabilistic uncertainty. The classic approach is to run a series of simulations with different parameter values randomly selected from the given probability distributions, which is referred to as the Monte Carlo (MC) method. The MC method has high computational cost, as the number of simulations required to accurately capture the probabilistic uncertainty is usually high, e.g., on the order of thousands (e.g., \cite{Kim2024BatteryPCE}), limiting its use in applications where fast or real-time computation is needed. As our framework is aimed for continuous manufacturing, the uncertainty analysis should be fast enough to implement in real time.

This article employs polynomial chaos theory (PCT) \cite{Wiener1938PCE}, which is a systematic, computationally efficient technique to propagate probabilistic uncertainty through the model equations. The key idea of PCT is to represent random variables with a set of orthogonal polynomials. This approach has been used for a range of applications including uncertainty quantification and stochastic control \cite{Mesbah2016SMPC,Matthias2019OffsetFreeSMPC}. 

\subsection{Mathematical formulation}
Consider the model consisting of $N_\theta$ independent random variables (inputs and parameters) represented by the vector $\boldsymbol{\Theta} = [\theta_1, \theta_2, \dots{}, \theta_{N_\theta}]^\top$. The output $Y$ approximated by the general truncated polynomial chaos expansion (PCE) of order $N_P$ is \cite{Eldred2008NonInPCE,Eldred2012NonInPCE,PoCET2020}
\begin{equation} \label{eq:PCE}
    Y = \sum_{i=0}^{L-1} y_i\psi_i(\boldsymbol{\Theta}) = \mathbf{y}^\top \boldsymbol{\Psi}(\boldsymbol{\Theta}),
\end{equation}
where $y_i$ is the PCE coefficient, $\psi_i$ is the $i^\textrm{th}$ multivariate polynomial basis, and $L = \frac{(N_\theta+N_P)!}{N_\theta! N_P!}$. The $i^\textrm{th}$ multivariate polynomial basis is a tensor product of the univariate polynomials, that is,
\begin{equation} \label{eq:basis}
    \psi_i\left(\boldsymbol{\Theta}\right) = \prod_{j=1}^{N_\theta}\varphi_{\nu_{i,j}}(\theta_j),
\end{equation}
where $\varphi_{\nu_{i,j}}$ is the univariate polynomial basis of order $\nu_{i,j}$ selected from the probability distribution of the variable $\theta_j$. The order of each multivariate polynomial basis $\psi_i$ is $\sum_{j=1}^{N_\theta}\nu_{i,j} \leq N_P$. A set of orthogonal polynomials that provide an optimal basis for various continuous probability distributions is given in Table \ref{Tab:Polynomials} \cite{Xiu2002Basis,Eldred2008NonInPCE,Eldred2012NonInPCE}.

\begin{table}[ht!]
\caption{Optimal basis functions in PCT.}
\vspace{-5pt}
\label{Tab:Polynomials}
\centering
\begin{tabular}{|c|c|c|}
\hline
Distribution & Polynomial & Support \\ 
\hline
Uniform & Legendre & $[-1,1]$  \\
Gaussian & Hermite & $(-\infty,\infty)$  \\
Beta & Jacobi &  $[-1,1]$  \\
Gamma  & Laguerre & $(0,\infty)$  \\
\hline
\end{tabular}
\end{table}

After the PCT-based model is formulated, the final step is to calculate the PCE coefficients. These coefficients can be calculated using either (1) intrusive or (2) non-intrusive approaches. The intrusive technique directly substitutes the PCE (e.g., \eqref{eq:PCE}) into the model equations, resulting in an expanded system of equations that is larger than the original, in which the PCE coefficients can be computed using the Galerkin projection \cite{Parekh2020CFD_IC,Son2020Compare}. In the non-intrusive technique, the original model equations and simulations are used as a black box to generate a set of responses, and so the PCE coefficients that best match the responses are calculated, e.g., using linear regression \cite{Eldred2008NonInPCE,Eldred2012NonInPCE}. Both approaches have been widely implemented in the literature. In this work, the non-intrusive approach is selected as its implementation is significantly simpler, particularly for nonlinear systems.

\subsection{Implementation}
Several tools that facilitate the implementation of PCE are currently available. We employed and modified PoCET \cite{PoCET2020} to generate a set of multivariate polynomials and compute the PCE coefficients. The number of sampling points was set to 2000 for MC and 50 for PCE, with $N_P = 2$. All calculations and simulations were performed in MATLAB 2023a on a computer equipped with an Intel\textsuperscript{\textregistered} Core\texttrademark\ i9-13950HX Processor CPU (24 cores) and 128 GB RAM on 64-bit Windows 11. 

\begin{figure*}[ht!]
\centering
    \vspace{5pt} 
    \includegraphics[scale=.92]{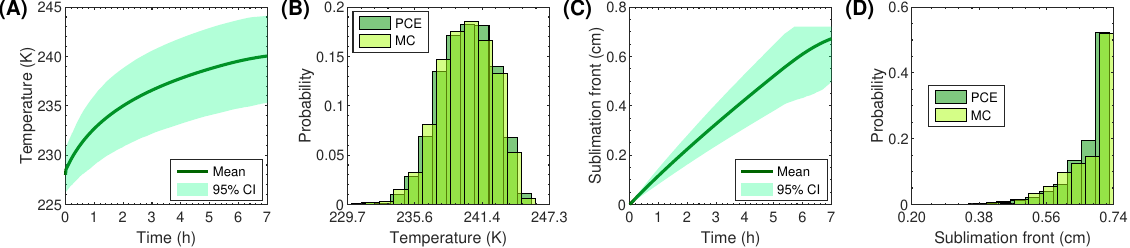}
    \vspace{-10pt}   
    \caption{Uncertainty quantification of the primary drying step. (A) Mean and 95\% confidence interval of the product temperature simulated via PCE. (B) Probability distribution of the temperature at the final time computed by PCE and MC. (C) Mean and 95\% confidence interval of the sublimation front simulated via PCE. (D) Probability distribution of the sublimation front at the final time computed by PCE and MC.} 
    \label{fig:PrimDrying_PCE_MC}      
\end{figure*}

\begin{figure*}[ht!]
\centering
    \includegraphics[scale=.92]{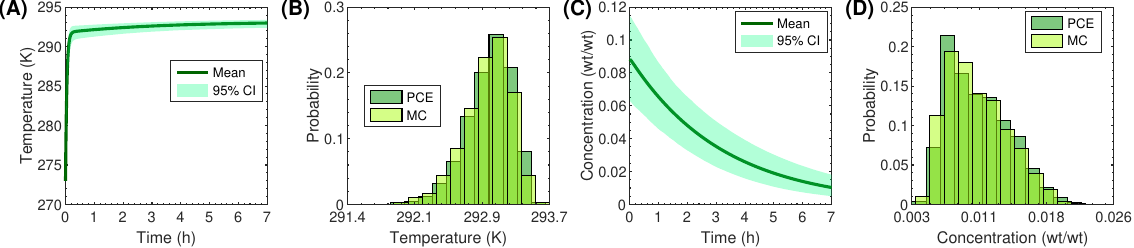}
    \vspace{-10pt}   
    \caption{Uncertainty quantification of the secondary drying step. (A) Mean and 95\% confidence interval of the product temperature simulated via PCE. (B) Probability distribution of the temperature at the final time computed by PCE and MC. (C) Mean and 95\% confidence interval of the concentration of bound water simulated via PCE. (D) Probability distribution of the concentration of bound water at the final time computed by PCE and MC.} 
    \vspace{-10pt} 
    \label{fig:SecDrying_PCE_MC}      
\end{figure*}

\section{Results and Discussion} \label{sec:Results}
This section presents case studies and simulation results associated with uncertainty quantification, optimization, and control of the continuous lyophilization process using the PCT-based model. Table \ref{Tab:Parameters} lists the uncertain parameters considered in this work; other parameter values and data can be found in our published software (see the Data Availability section) and \cite{Srisuma2024ContModeling}.

\begin{table}
\vspace{5pt}
\caption{Uncertain parameters.} \label{Tab:Parameters}
\vspace{-7pt}
\centering
\begin{threeparttable}
\begin{tabular}{|c|c|c|}
\hline
Symbol & Value/Distribution & Unit \\ 
\hline
$h$ & Gaussian, $\mu=15,\sigma=3$ &  W/(m$^2$$\cdot$K)  \\
$R_0$ & Uniform, [1$\times$$10^{4}$, 2$\times$$10^{4}$] & m/s  \\
$R_1$ & Uniform, [1$\times$$10^{7}$, 3$\times$$10^{7}$] & 1/s  \\
$f_\textrm{a}$ & Uniform, [0.3, 0.5] & 1/s  \\
$c_\textrm{w,0}$ &  Gaussian, $\mu=0.088,\sigma=0.018$ & wt/wt\tnote{a}  \\
\hline
\end{tabular}
\begin{tablenotes}
{\footnotesize \item[a]wt/wt refers to kg water/kg solid.}
\end{tablenotes}
\end{threeparttable}
\vspace{-15pt}
\end{table}

\subsection{Uncertainty quantification} \label{sec:UQ}
First consider the primary drying step, denoted as Case A1. Common uncertain parameters in primary drying are $R_\textrm{p}$ and $h$ \cite{Giordano2011Model}. The cake resistance $R_\textrm{p}$ is influenced by the freezing step, which is inherently a stochastic process, whereas the heat transfer coefficient $h$ is affected by various factors, e.g., material properties and operating conditions.

\begin{figure*}[ht!]
\centering
    \vspace{5pt}  
    \includegraphics[scale=.9]{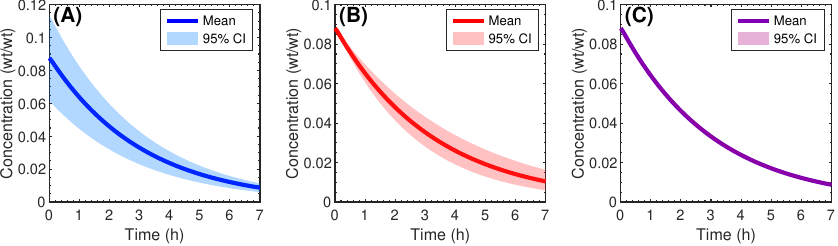}
    \vspace{-9pt}   
    \caption{Evolution of the concentration of bound water for the uncertain (A) initial concentration, (B) desorption kinetics, and (C) heat transfer coefficient.}
    \vspace{-10pt}  
\label{fig:SecDrying_PCE_All}    
\end{figure*}

The effects of uncertainty are quantified for the product temperature (Fig.~\ref{fig:PrimDrying_PCE_MC}AB) and sublimation front (Fig.~\ref{fig:PrimDrying_PCE_MC}CD). The product temperature increases from the initial value of 228 K to about 240 K at the end of the process (Fig.~\ref{fig:PrimDrying_PCE_MC}A). The final temperature could vary between 230 K to 247 K (Fig.~\ref{fig:PrimDrying_PCE_MC}B). The effect of uncertainty on the sublimation front is relatively weak at the beginning of the process and becomes stronger after 1 h (Fig.~\ref{fig:PrimDrying_PCE_MC}C). The probability distribution predicted by PCE closely aligns with that computed by MC in both cases (Fig.~\ref{fig:PrimDrying_PCE_MC}BD).


Next consider the secondary drying step, denoted as Case A2. Typical uncertain parameters are $c_\textrm{w,0}$, $k_\textrm{d}$, and $h$. The initial concentration $c_\textrm{w,0}$ could vary among different freezing and primary drying conditions, and hence it is not always known precisely. The desorption kinetics $k_\textrm{d}$ is dependent on various factors, e.g., surface properties and solid structure.  

The effects of uncertainty are quantified for the product temperature (Fig.~\ref{fig:SecDrying_PCE_MC}AB) and concentration of bound water (Fig.~\ref{fig:SecDrying_PCE_MC}CD). The product temperature is nearly insensitive to the uncertainty as it approaches the shelf temperature relatively quickly (Fig.~\ref{fig:SecDrying_PCE_MC}A), which is because the supplied heat is entirely used for increasing the temperature, not for sublimation as in primary drying. The concentration of bound water, on the other hand, is significantly influenced by the uncertainty, with the 95\% confidence interval (CI) of about 0.02 to 0.05 wt/wt throughout the process (Fig.~\ref{fig:SecDrying_PCE_MC}C). The probability distribution predicted by PCE agrees well with the result obtained from MC for both temperature and concentration (Fig.~\ref{fig:SecDrying_PCE_MC}BD). 


The effects of all uncertain parameters are considered simultaneously in Fig.~\ref{fig:SecDrying_PCE_MC}. To provide more insights into the secondary drying case, the effect of each uncertain parameter is investigated individually (Fig.~\ref{fig:SecDrying_PCE_All}). The uncertain initial concentration contributes significantly at the beginning of the process but almost has no effect at the end, e.g., after 5 h (Fig.~\ref{fig:SecDrying_PCE_All}A). On the other hand, the uncertain desorption kinetics greatly impacts the final concentration (Fig.~\ref{fig:SecDrying_PCE_All}B). This level of uncertainty in heat transfer appears to produce negligible effects on the concentration profile in this case (Fig.~\ref{fig:SecDrying_PCE_All}C).


\subsection{Model-based design and optimal control} \label{sec:Control}
The PCT-based mechanistic model is demonstrated above to provide reliable uncertainty quantification. This section extends that capability for process design, optimization, and control. Here we focus more on secondary drying as it receives much less attention than primary drying but can also greatly impact the final product quality.

Case B1 employs the model to design a continuous lyophilization process. The objective is to find the minimum shelf temperature required to achieve a target drying time, with the final concentration of bound water (average spatially) that is below the target value within the acceptable probability, denoted as $\mathbb{P}$. A typical target concentration is 1\% or 0.01 wt/wt \cite{Fissore2018Review}. For this demonstration, $\mathbb{P} = 0.95$ and the target drying time is set to 7 h.

\begin{figure}[ht!]
\centering
\vspace{6pt}
\includegraphics[scale=.85]{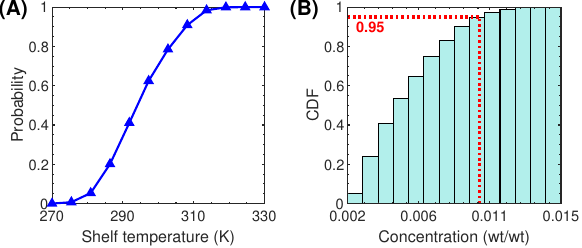}
    \vspace{-5pt} 
    \caption{(A) Probability that the final concentration of bound water is below 0.01 wt/wt for different shelf temperatures. (B) Cumulative distribution function (CDF) at the optimal shelf temperature of 310 K.} 
    \vspace{-13pt} 
    \label{fig:SecDrying_Opt2A}
\end{figure}

A series of simulations with different values of $T_b$ is investigated. The probability that the final concentration meets the target increases with an increase in the shelf temperature (Fig.~\ref{fig:SecDrying_Opt2A}A), which is reasonable as desorption is faster at higher temperature. The minimum shelf temperature required to achieve the target drying time and probability is about 310 K (Fig.~\ref{fig:SecDrying_Opt2A}B).

\begin{figure}[ht!]
\centering
    \vspace{6pt} 
    \includegraphics[scale=.85]{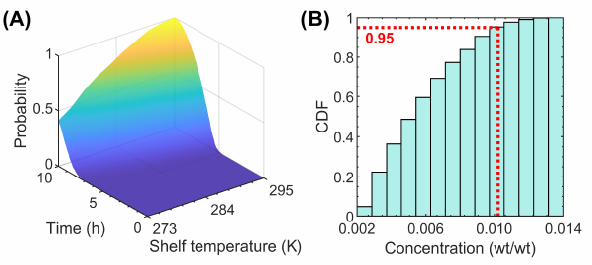}
    \vspace{-5pt} 
    \caption{(A) Probability that the final concentration of bound water is below 0.01 wt/wt for different drying times and shelf temperatures. (B) Cumulative distribution function (CDF) at the optimal drying time of 9.0 h and shelf temperature of 295 K.} 
    \vspace{-15pt} 
    \label{fig:SecDrying_Opt2B}      
\end{figure}

Case B2 considers an optimal control problem, in which the objective is to minimize the total drying time. Besides, the final concentration needs to satisfy the probabilistic constraint defined in Case B1, leading to the optimization 
\begingroup
\allowdisplaybreaks
\begin{equation} \label{eq:OptB1}
\begin{aligned} 
&\min_{T_b(t)} \ t_f \\
& \  \textrm{s.t.} 
    \ \textrm{\eqref{eq:2nd_energy}--\eqref{eq:2nd_inicw}}, \\
   & \ \textrm{273 K} \leq T_b(t) \leq \textrm{295 K},\\
& \ \mathbb{P}(c_{\textrm{w}}(t_f)\leq0.01 \ \textrm{wt/wt}) = 0.95, 
\end{aligned}
\end{equation}
\endgroup
where $t_f$ is the total drying time. Physically, an increase in the shelf temperature should accelerate the drying process (as observed in Case B1), and thus, without any further constraints, the shelf temperature should be at its upper bound. Using MATLAB's \texttt{fmincon} to solve \eqref{eq:OptB1} gives the optimal drying time of about 9.0 h and shelf temperature of 295 K\footnote{Under a steady-state operation of continuous lyophilization, the shelf temperature is constant, and the control vector parameterization can be carried out with 1 interval (degree of freedom); the same procedure is applied for systems with higher degrees of freedom, but with higher order parameterizations of the control trajectory.} (Fig.~\ref{fig:SecDrying_Opt2B}). The optimal shelf temperature agrees with our expectation, with the chance constraint satisfied properly.


\subsection{Computational performance} \label{sec:Performance}
Finally, to demonstrate the benefit of our PCT-based model over the traditional MC approach, this section compares the computation time used by the PCE and MC methods for all the case studies in Sections \ref{sec:UQ} and \ref{sec:Control}. 

The computation (wall-clock) times measured with the \texttt{tic} and \texttt{toc} functions in MATLAB are given in Table \ref{Tab:ComputationalPerformance}. The computation time required for running a single simulation with PCE (Cases A1 and A2) is less than 1 s, which is much faster than for the MC method. This computational benefit is even more obvious when a large number of simulations is needed, e.g., design optimization and optimal control as in Cases B1 and B2. Overall, PCE is a much more computationally efficient approach, which is at least an order of magnitude faster than MC.

\section{Conclusion} \label{sec:Conclusion}
This work presents a framework to explicitly account for the probabilistic uncertainty during the primary drying and secondary drying steps in continuous lyophilization of suspended vials. A PCT-based mechanistic model is developed and validated with the Monte Carlo method. The resulting PCT-based model  accurately and efficiently quantifies the effects of uncertainty on the product temperature, sublimation front, and concentration of bound water. The cake resistance and heat transfer coefficient are critical parameters in primary drying, whereas the desorption kinetics and initial concentration are found to be important in secondary drying.  Besides uncertainty quantification, the model is demonstrated for process design and optimization under uncertainty, in which the final concentration is well controlled within the target value at a given probability.

\begin{table}[ht!]
\caption{Computational performance of the PCE and MC methods from 10 simulation runs of each case study.}
\vspace{-5pt}
\label{Tab:ComputationalPerformance}
\centering
\begin{tabular}{| c |  wc{8em} |  wc{8em} |}
\hline
\multirow{2}{*}{Case} & \multicolumn{2}{c|}{Computation time (s)} \\ \cline{2-3}
&  PCE & MC \\
\hline
A1 &  $0.56\pm0.01$ & $13.98\pm0.10$ \\
A2 & $0.47\pm0.01$ & $5.24\pm0.12$  \\
B1 & $4.02\pm0.10$  & $85.99\pm2.80$  \\
B2 & $2.86\pm0.04$ & $64.96\pm0.63$  \\
\hline
\end{tabular}
\end{table}
\vspace{-7pt}

\section*{Data Availability}  \label{sec:Code}
Software and data used in this work are available at \url{https://github.com/PrakitrSrisuma/ContLyo-PCE-drying}.

\section*{Acknowledgments} 
This research was supported by the U.S. Food and Drug Administration under the FDA BAA-22-00123 program, Award Number 75F40122C00200.


\bibliographystyle{IEEEtran}
\bibliography{IEEEabrv,reference}

\end{document}